\documentclass[sigconf,nonacm=true,natbib=false,pbalance=true,urlbreakonhyphens=true]{acmart} %

\usepackage{etoolbox} %

\usepackage[abbreviate=true, dateabbrev=true, alldates=year, isbn=false, doi=false, urldate=comp, url=true, maxbibnames=9, backref=false, backend=biber, style=numeric-comp, language=american, giveninits=true, sortcites=true, sorting=none]{biblatex}
\AtEveryBibitem{\clearlist{location}} %
\AtEveryBibitem{\clearfield{series}} %
\AtEveryBibitem{\clearlist{language}} %

\copyrightyear{2022}
\acmYear{2022}
\setcopyright{acmcopyright}
\acmConference[ICSE '22 Companion]{44th International Conference on Software Engineering Companion}{May 21--29, 2022}{Pittsburgh, PA, USA}
\acmBooktitle{44th International Conference on Software Engineering Companion (ICSE '22 Companion), May 21--29, 2022, Pittsburgh, PA, USA}
\acmPrice{15.00}
\acmDOI{10.1145/3510454.3517063}
\acmISBN{978-1-4503-9223-5/22/05}

\newtoggle{arxiv}
\toggletrue{arxiv}
\iftoggle{arxiv}{%
  \thanks{\raisebox{21.5pt}{\hbox to 0.951\columnwidth{\parbox{0.99\columnwidth}{\small This is a preprint of the work published in the 44th International Conference on Software Engineering 
    (ICSE ’22) Doctoral Symposium, May 21–29, 2022, Pittsburgh, PA, USA, \url{https://doi.org/10.1145/3510454.3517063}. \\[1ex]%
    \includegraphics[width=2.3cm]{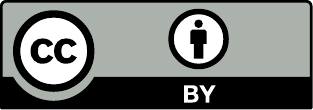}
      \raisebox{9pt}{\hspace*{2mm}\parbox{5.7cm}{\small This version is licensed under a 
       \href{https://creativecommons.org/licenses/by/4.0/}{Creative Commons Attribution 4.0 International (CC BY 4.0)} license}}}}}}%
}

\acmSubmissionID{DS-11}

\begin{document}

\title{Enabling Automatic Repair of Source Code Vulnerabilities Using Data-Driven Methods}

\author{Anastasiia Grishina}
\orcid{0000-0003-3139-0200}
\affiliation{%
  \institution{Simula Research Laboratory}
  \streetaddress{Kristian Augusts gate 23}
  \city{Oslo}
  \country{Norway}
  \postcode{0164}
}
\email{anastasiia@simula.no}

\renewcommand{\shortauthors}{A. Grishina}

\begin{abstract}
Users around the world rely on software-intensive systems in their day-to-day activities. 
These systems regularly contain bugs and security vulnerabilities.
To facilitate bug fixing, data-driven models of automatic program repair use pairs of buggy and fixed code to learn transformations that fix errors in code. 
However, automatic repair of security vulnerabilities remains under-explored. 
In this work, we propose ways to improve code representations for vulnerability repair from three perspectives: input data type, data-driven models, and downstream tasks. 
The expected results of this work are improved code representations for automatic program repair and, specifically, fixing security vulnerabilities.
\end{abstract}

\keywords{
Automatic Program Repair, 
Static Analysis,
Software Security, 
Natural Language Processing, 
Graph-based Machine Learning,
ML4Code}

\maketitle

\section{Introduction}

Users around the world rely on various software-intensive systems, such as the systems that assist 
in healthcare and transportation. 
These systems regularly contain bugs and security vulnerabilities that can have various negative effects, from economical to life-threatening~\cite{bbvaopenmind2015:most, wilson2016:bugs}. 
Software companies have been making a considerable effort to eliminate such bugs and vulnerabilities. 
To support these activities, Automatic Program Repair (APR) methods have received increasing attention in software engineering research~\cite{monperrus2018:automatic}.

The rich body of literature on APR has been surveyed in various studies~\cite{allamanis2018:survey, liuyuzhen2018:survey, monperrus2018:automatic, gazzola2019:automatic, goues2019:automated, sharma2021:survey}. 
Following the terminology of the APR surveys~\cite{monperrus2018:automatic, goues2019:automated}, this PhD project focuses on 
\emph{learning-based methods} of \emph{static} repair of \emph{source code vulnerabilities}. 
In \emph{static} analysis, we process code without executing it, as opposed to dynamic analysis. 
When approached with \emph{learning-based methods}, APR requires training data. 
We also refer to these methods as \emph{data-driven methods}.
Using code 
organized as pairs of (buggy code; fixed code), data-driven approaches employ deep learning techniques to 
predict a fix, also referred to as a patch.

The main objective of this work is to improve automatic vulnerability repair in source code from the following three perspectives. 

(a) \emph{Input data}. Currently, code change history is the major source of input data for training code representation models. 
In addition, forum discussions are used jointly with code change history for code analysis and repair~\cite{liu2018:mining, mahajan2020:recommending, islam2020:repairing}. 
This work will explore the effect of integrating context from additional input modalities, such as programming tutorials, on security-oriented APR.

(b) \emph{Models}. Two characteristics of source code are used in literature: its similarity to natural languages and the formal structure of code.
In the modelling phase, this work will focus on creating hybrid methods based on these two approaches for code representation. 

(c) \emph{Downstream tasks}. To guarantee the security of given software, one approach is to eliminate the known vulnerabilities.
Instances of security principles violation are published in Common Vulnerabilities and Exposures (CVE) Records.\footnote{~\url{https://cve.mitre.org/}}
However, the CVE list is not exhaustive, and not all CVEs are bugs.
For example, issues, such as
storing passwords without salt or using an unreliable hashing function, are commonly not reported as errors by compilers, but they can compromise confidentiality and integrity of an application. 
Moreover, repair of security vulnerabilities is under-explored, compared to general bugs~\cite{sharma2021:survey}.
Therefore, we have chosen pattern mining for CVE types and vulnerability repair as downstream tasks.

\section{Related Work}

Two strands of research in data-driven APR methods form the basis for this work. 
They consider different approaches to code representation that are used for various downstream tasks, including code repair. 
One branch focuses on similarities between natural language and code, while the other
represents code using its structure. 

Allamanis et al.~\cite{allamanis2018:survey} presents the hypothesis that programming languages are forms of natural language. Therefore, Natural Language Processing (NLP) can be reused for code analysis tasks. %
Examples of code processing models are Tufano et al.~\cite{tufano2019:empirical}, SequenceR~\cite{chen2019:sequencer}, CodeBERT~\cite{feng2020:codebert}, Codex~\cite{chen2021:evaluating}. 
These models reuse RNNs~\cite{sutskever2014:sequence}, transformer architectures~\cite{devlin2019:bert} with attention~\cite{vaswani2017:attention}, and the copy mechanism~\cite{gu2016:incorporating}. 
Specifically, CodeBERT is a recent study on fine-tuning the BERT model on a large varied code base to perform tasks related to translating from natural language to programming language and the same in the reverse direction. %
Another study that reused BERT is CuBERT~\cite{kanade2020:learning}. The authors trained the BERT model on code and fine-tuned it on code understanding and repair tasks. 
A recent advancement in code representation by GitHub and OpenAI, Codex~\cite{chen2021:evaluating} empowering Copilot, is built using the GPT-3 model. Codex translates a human language description of a 
function to code performing the described task.

One of the differences between source code and natural language is strict syntactic and logical structure of code that can be used for code representation. 
Code parsed according to its syntax forms an Abstract Syntax Tree (AST)~\cite{long2017:automatic}.
Similarly, data flow and control flow graphs developed for code analysis earlier also represent code characteristics~\cite{stevens1974:structured, allen1970:control}. 
For instance, code2vec~\cite{alon2019:code2vec} uses both n-grams (n words considered as one token) and ASTs to automatically predict function name from its code. 
GraphCodeBERT~\cite{guo2021:graphcodebert} obtains an AST and translates it to a data flow graph to obtain one of three input representations of code that are tested on four downstream tasks, including code repair.
Although they have received less attention than the mentioned graph types, knowledge graphs are also used in code analysis and repair tasks~\cite{atzeni2017:codeontology, nguyen2019:graphbased, djebri2021:linked}.

While the aforementioned research focuses on general program repair, we concentrate on repair of security vulnerabilities. Compared to security vulnerability detection and general bug repair, security vulnerability repair has received less attention~\cite{shen2020:survey, sharma2021:survey}. For example, the benchmark CodeXGlue~\cite{lu2021:codexglue} has a setup for vulnerability detection, but the repair (refinement) part contains only general bugs following the dataset of Tufano et al.~\cite{tufano2019:empirical}.
One study on vulnerability repair uses generative adversarial networks to generate security vulnerability patches in C/C++~\cite{harer2018:learning}. Another study focuses on repair of vulnerabilities of code generated by Codex~\cite{pearce2021:can}.
In this research, we aim at improving results for vulnerability repair and generalize to other programming languages.
\vspace{-2pt}

\section{Expected Contributions}

\begin{figure}[t]
	\centering
	\includegraphics[width=0.48\textwidth, trim=0 6 0 0, clip]{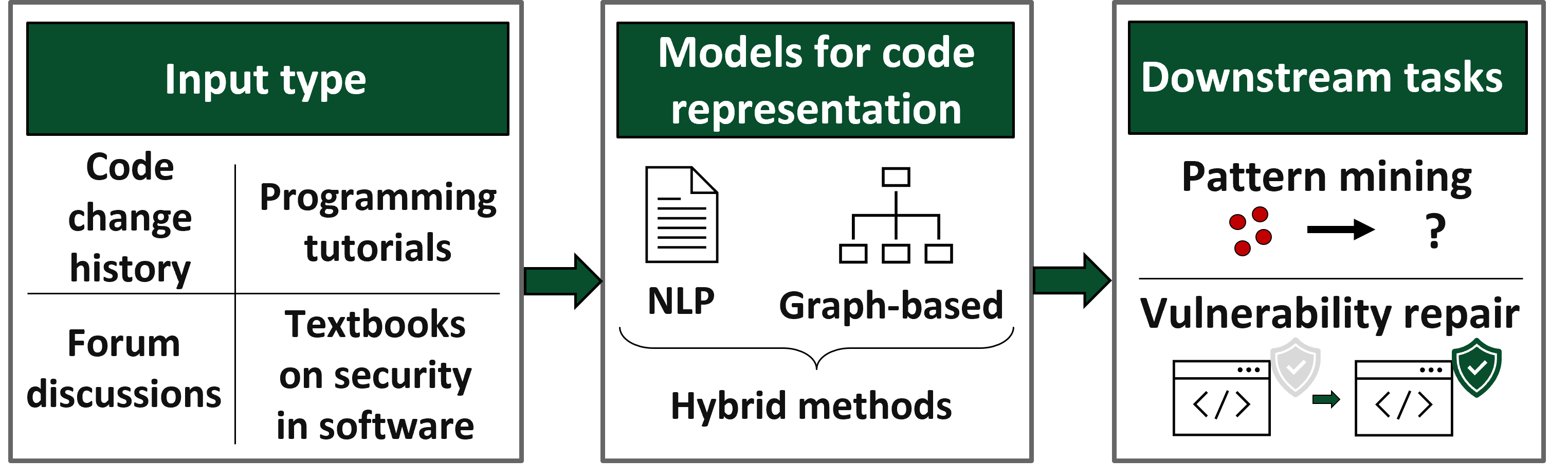}%
 	\vspace*{-2ex}
	\caption{Organization of the work: we focus on (a) different input modalities; (b) model types;
    (c) two downstream tasks.}
	\label{fig:overview}
	\vspace{1.5ex}
\end{figure}

The overview of this work is sketched in Figure~\ref{fig:overview}. In the three phases of the PhD project, 
we hypothesize that 
(a) using context from additional modalities can improve APR performance; 
(b) when built upon NLP-based and graph-based methods for code representation and generation, hybrid methods can capture both natural language similarities and structural properties of code; 
(c) mining patterns from code in the wild can help 
improve vulnerability repair results. 
The overall objective is to automatically repair software vulnerabilities in source code using data-driven methods. 

Firstly, we aim to explore the impact of context from additional modalities on code representation and bug repair compared to the mostly exclusive use of code change history. 
Examples of input modalities are texts and code snippets in tutorials, books on security in software engineering, and forum discussions. 

Secondly, 
we aim to analyze to what extent existing APR models can be adapted to repair vulnerabilities. 
As baselines, we will use NLP-based and graph-based methods. 
In this step, we will experiment with building hybrid methods using baseline approaches. 
To the best of our knowledge, there is no work on combination of these methods for vulnerability repair. 

Finally, we will use code representations to mine patterns of errors and their fixes within each CVE type. The pattern mining task is inspired by the study that categorizes common mistakes in AI projects~\cite{islam2020:repairing}.
With patterns discovered for security vulnerability types, we will focus on vulnerability repair in source code. 

\section{Planned Evaluation of Results}

In this project, we will reuse data gathered in CVEfixes~\cite{bhandari2021:cvefixes}. CVEfixes provides both a dataset and a framework for collecting data from GitHub, GitLab, and Bitbucket. 
The already extracted dataset refers to CVE Records in the U.S. National Vulnerability Database to classify code snippets by CVE type. 
In this dataset, code changes are extracted on the file and method level. 
In downstream tasks, CVE type-specific pattern mining will use vulnerable code classified by CVE type, while code pairs of (vulnerable code; fixed code) will be used for training and evaluation of vulnerability repair.

The target for the modelling phase (b) is to learn vector representations of code as sequences of tokens or as graphs, so that code snippets of the same CVE type in the same programming language are close to each other in the embedding space. 
To measure the quality of code representations, we will use metrics for clustering in machine learning and metrics in NLP: Silhouette Coefficient, Normalized Mutual Information, Completeness, and V-measure for clusters; perplexity, cosine and other similarity metrics from NLP. 

Alternatively, since APR is analogical to the NLP task of neural machine translation, it can be evaluated with the Recall-Oriented Understudy for Gisting Evaluation (ROUGE) and Bilingual Evaluation Understudy (BLEU) NLP metrics~\cite{feng2020:codebert, clement2020:pymt5}, and their extensions~\cite{ren2020:codebleu, lu2021:codexglue}. 
In the context of vulnerability repair, ROUGE scores evaluate the patch based on the number of occurrences of n-grams from the known repaired code (reference sequence) in the patch (generated sequence). 
By contrast, BLEU shows n-gram precision of the sequence generation.
In detail, it is measured based on the number n-gram occurrences from the patch (generated sequence) that are also found in the known repaired code (reference sequence).

\vspace*{-2pt}
\section{Conclusion}

In conclusion, this project focuses on automatic repair of source code vulnerabilities. We hypothesize that repair performance can be improved with the context extracted from different modalities in addition to widely used code change history. Furthermore, the repair performance can be influenced by combining models that build on naturalness hypothesis and formal characteristics of code, such as syntax, structure, and dependencies. We aim to test methods of pattern mining for each vulnerability type, and its repair. Training will be done using the CVEfixes dataset, and the evaluation step will reuse 
clustering and NLP metrics. The expected results of this work are improved code representations for APR and, specifically, fixing security vulnerabilities.

\vspace*{-2pt}
\begin{acks}
This work is supported by the Research
Council of Norway through the secureIT project (IKTPLUSS \#288787).
\end{acks}

\sloppy %
\printbibliography

\end{document}